# Beyond 300Gbps Silicon Microring Modulator with AI Acceleration

**FANGCHEN HU,**[1,5] **YUGUANG ZHANG,**[3,5] **HONGGUANG ZHANG,**[3,5] **ZHONGYA LI,**[1] **SIZHE XING,**[1] **JIANYANG SHI,**[1] **JUNWEN ZHANG,**[1,*] **XI XIAO,**[2,3,4*] **NAN CHI,**[1,*] **ZHIXUE HE,**[2,4] AND **SHAOHUA YU**[2,3,4]

[1] *Key Laboratory for Information Science of Electromagnetic Waves (MoE), Fudan University, Shanghai 200433, China*
[2] *State Key Laboratory of Optical Communication Technologies and Networks, China Information Communication Technologies Group Corporation, Wuhan, Hubei 430074, China*
[3] *National Information Optoelectronics Innovation Center, 430074 Wuhan, China*
[4] *Peng Cheng Laboratory, Shenzhen, Guangdong 518055, China*
[5] *These authors contributed equally.*
*\* junwenzhang@fudan.edu.cn, xxiao@wri.com.cn, nanchi@fudan.edu.cn*

**Abstract:**

Silicon microring modulator (Si-MRM) has become one of the most promising compact modulators to meet the increasing capacity requirements of the next generation optical interconnection. The limited electro-optical (E-O) bandwidth, low modulation efficiency, and inherent modulation nonlinearity are the major factors that limit the Si-MRM modulation speed. To address these issues, we comprehensively optimize the Si-MRM from the device to the modulation and the signal processing. Large modulation bandwidth over 67GHz is achieved in our newly fabricated Si-MRM. Additionally, the laser wavelength and bias voltage of Si-MRM are optimized to significantly improve the modulation performance. Finally, we comprehensively study the theoretical model of modulation nonlinearity in Si-MRM, especially transient nonlinearity. A bidirectional gate recurrent unit (Bi-GRU) neural network with minor modification is applied to compensate for the nonlinear impairments. With all these efforts, we experimentally demonstrate a 302 Gbps Si-MRM-based O-band optical interconnection and achieve 300 Gbps transmission over a 1-km standard single-mode fiber using the discrete multitone modulation format with bit and power loading (BPL-DMT).

## Introduction

Silicon photonics and optical microresonators have enabled a wide range of significant applications in the fields of quantum electrodynamics, sensing, optical telecommunication and photonic integrated circuits [1-4]. One of the most critical components in these applications is the silicon microring modulator (Si-MRM) which serves for a high-performance electrical signal to optical signal conversion. It has become a promising candidate to meet next-generation high-speed optical interconnections [5, 6] due to the merits of ultrawide bandwidth, low loss, and complementary metal-oxide-semiconductor (CMOS) compatibility [7, 8]. The Si-MRM also has a micron-sized footprint and simple driver structures and thus has lower power consumption than other silicon modulators, such as the one based on a Mach-Zehnder interferometer. These advantages make it perfectly suitable for applications where the power, area, and cost are highly constrained and require high communication capacity [9, 10].

Several challenges must be overcome to realize a high-capacity Si-MRM-based optical interconnection. A significant challenge is to achieve an ultrawide electrooptical (E-O) modulation bandwidth for Si-MRMs. In addition to the essential method of optimizing the resistance and capacitance of P-N junctions, optical peaking enhancement (OPE) technology has emerged as an appealing option to further extend the optical bandwidth by the dynamics of the Si-MRM [11]. The achievable bandwidth gain depends on the wavelength detuning, defined as the difference between the laser and resonance wavelengths. Many Si-MRMs have demonstrated modulation capacities beyond 100 Gbps with E-O bandwidths over 50 GHz due to OPE technology [12-14]. However, there exists an intrinsic trade-off between the bandwidth and modulation efficiency due to the resonance of Si-MRMs [15]. Specifically, the improvement of the bandwidth by OPE technology comes at the cost of the overall optical modulation amplitude. In addition, the signal waveform suffers from more complex nonlinear distortions, including a time-domain peak induced by the OPE technology observed in several experimental studies [16-18]. These nonlinear impairments on signal eventually lead to decreased modulation efficiency. Therefore, another major challenge is how to mitigate these nonlinear impairments to maximize the modulation efficiency while preserving the bandwidth gain from OPE technology.

Nonlinear impairments induced by the modulation nonlinearity of Si-MRM can be categorized into static nonlinear impairments [19] and transient nonlinear impairments. Besides the different frequency response for different voltages [20], the transient nonlinear impairment also results in the overshooting phenomenon. The overshooting phenomenon was observed for the first time by Q. Xu [16]. Q. Xu inferred from the physics perspective that the overshoot arises from the interference of the trapped light in the ring resonator with the transmitted light when the trapped light couples back to the bus waveguide due to fast wavelength detuning. Based on this inference, L. Zhang proved that the overshoot phenomenon widely exists in both undercoupled and overcoupled Si-MRMs, related to the frequency chirp. Overall, studies on transient nonlinear impairment remain in a preliminary state with sporadic analytical thinking, lacking a complete theoretical analysis of its mechanism and dependency, which inversely is most instructive for a targeted equalizer design. Here, based on the dynamic characteristics of Si-MRMs, we theoretically analyze the transient nonlinear impairment and prove that it has a time-relevant pattern dependency on the signal levels.

Recent advances in AI have substantially impacted many scientific and engineering fields including medicine, material inspection, climates prediction and bioinformatics, etc [21]. The AI technology could use its powerful computational modeling ability to obtain intelligent solutions for some problems that virtually impossible to explicitly formulate before [22]. In optics, AI successfully enhances the optics design optimization [23], optical image reconstruction [24] and resolution of microscopy [25]. AI is also applied to combat nonlinear distortions in optical communication [26-29]. A long short-term memory (LSTM) equalizer, one variant of recurrent neural networks, is believed to mitigate fiber dispersion effectively and the bandwidth limit of the Si-MRM-based communication system reported in Ref. [30]. However, this paper neglects LSTM also mitigates the nonlinear impairments from Si-MRM. The poor explanation of why and how LSTM could improve Si-MRM-based transmission makes its performance untrusted and uncontrolled for users, although the neural network always exhibits unparalleled performance [31]. The underlying mathematical models remain elusive to interpretation by the human mind. The selection of neural networks should approach the theoretical model of the physics problem, which fits the interpretation on equalization mechanism by the human mind. Such a neural network can provide efficient equalization performance with high confidence.

In this paper, we aim to address challenges mentioned above, and demonstrate the AI-accelerated ultra-high-speed optical interconnection based on our newly fabricated ultra-high-bandwidth Si-MRM. We first fabricate an ultrahigh-bandwidth Si-MRM with a -3-dB optical bandwidth beyond 67 GHz thanks to the doping concentration optimization and OPE technology. In addition, the laser wavelength and bias voltage of electrical signal are optimized to improve the overall modulation capacity. The theoretical model of modulation nonlinearity in Si-MRM, especially transient nonlinearity is comprehensively studied. A bidirectional gate recurrent unit (Bi-GRU) neural network with minor modification based on the obtained physical model of Si-MRM's modulation nonlinearity is applied. The inner configuration of Bi-GRU matches the dynamic characteristics of Si-MRM, which helps to effectively capture features of the nonlinear impairments of Si-MRM. The Bi-GRU neural network provides a more than 1.1-dB received optical power gain than traditional nonlinear equalizers. With all these efforts, we demonstrate a 302 Gbps O-band Si-MRM-based optical interconnection and 300 Gbps transmission over a 1-km standard single-mode fiber using the discrete multitone modulation format with bit and power loading (BPL-DMT). To the best of our knowledge, a 302 Gbps modulation capacity is the highest data rate ever reported for the Si-MRM. The results enhance the understanding of Si-MRM's high-speed modulation characteristics, meanwhile confirming the significant value of AI technologies in optical communication.

**Results**

*Design of the Si-MRM*

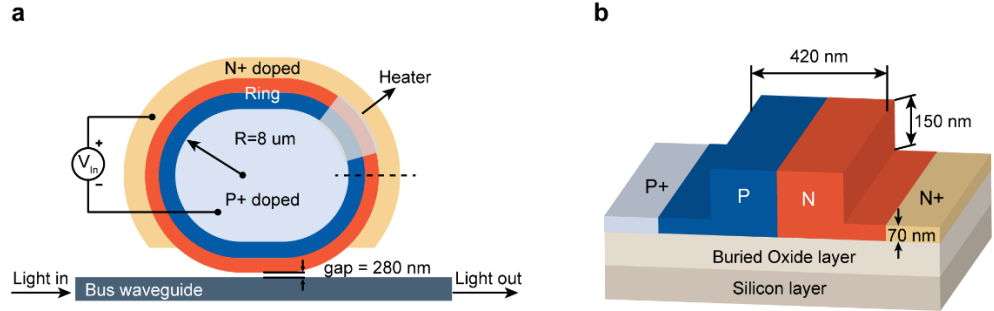

**Fig. 1 Schematic of a Si-MRM.** (a) Top view: A P-N diode is embedded in the racetrack ring resonator. The black dashed line indicates the position of the (b) 3-D cross-sectional view.

A schematic of the proposed Si-MRM is depicted in Fig. 1. It consists of a racetrack ring resonator and a bus waveguide. Light to be transmitted first enters one side of the bus waveguide and then travels around the ring resonator before being output from the other side of the bus waveguide. If a reverse voltage is imposed on the P-N junction embedded in the ring resonator, the intensity of the electrical signal could be modulated on the effective index of the ring waveguide, thereby affecting the optical power of the output light through the free-carrier dispersion effect [33]. The electrical modulation properties of the Si-MRM critically rely on the design of the ring resonator. The radius (R) of the ring resonator is 8 μm to make the resonance wavelength work at the O-band (approximately 1310 nm), in which the measured free spectrum range (FSR) is 7 nm. Since the resonance wavelength is also sensitive to temperature, a TiN heater is mounted on the ring resonator that can further accurately control the resonance wavelength. The temperature-optics (T-O) efficiency is 0.1 nm/mW. In addition, the resistance and capacitance of the Si-MRM are well optimized to extend the optical bandwidth beyond 67 GHz with a detuning frequency of 32 GHz. Optical peaking appears at the frequency point of approximately 50 GHz. The product of the π-phase-shift voltage and length remains as high as 0.8 V·cm. More characteristics of the Si-MRM are reported in Ref. [34].

According to the standard CMOS process, the Si-MRM is fabricated on a silicon-on-insulator (SOI) platform with a 2-μm-thick buried oxide layer. The ring waveguide is obtained by etching a 220-nm width Si waveguide with an etching depth of 150 nm. A cross-sectional view of this slab waveguide is illustrated in Fig. 1b. Both it and the bus waveguide have a width of 420 nm to reduce the propagation loss and to meet the single-mode condition. A P-N junction is embedded in the ring resonator with an optimized doping concentration of $3\times10^{18}cm^{-3}$. A specifically designed racetrack with a length of 5 μm is placed between the ring waveguide and the bus waveguide with a gap of 280 nm to ensure a high coupling efficiency.

*Theoretical analysis of the nonlinear impairment in a Si-MRM*

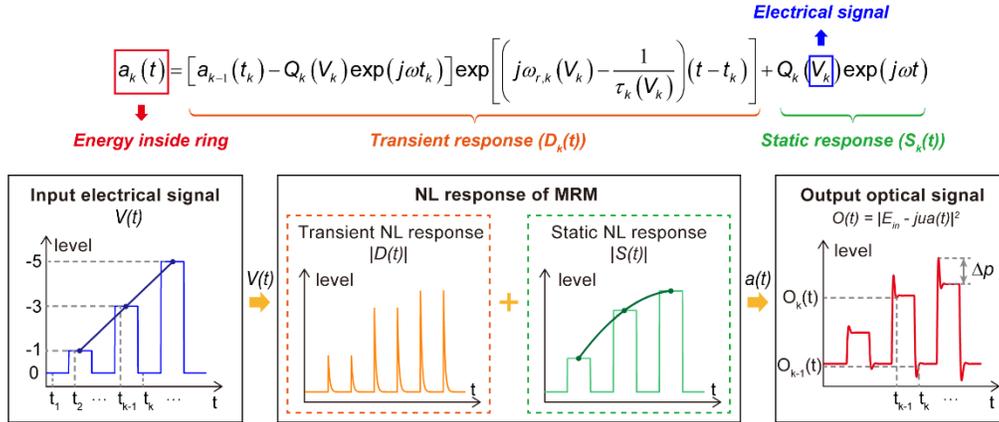

**Fig. 2 Theoretical model and modulation nonlinearity of the Si-MRM for electrical intensity modulation.** The iteration equation of energy stored inside the ring resonator in the $k_{th}$ time slot is given ($t_{k-1} \leq t \leq t_k$). The waveform of the input multilevel electrical signal, the optical signal output from the Si-MRM and the nonlinear (NL) distortion induced by transient and static responses of the Si-MRM are shown.

The dynamic characteristics of the Si-MRM are always described by coupled mode equations [35]:

$$\frac{d}{dt}a(t) = \left(j\omega_{res} - \frac{1}{\tau}\right)a(t) - j\mu E_{in} \tag{1}$$

$$E_{out}(t) = E_{in}(t) - j\mu a(t) \tag{2}$$

where $E_{in}$ and $E_{out}$ represent the incident and output light of the Si-MRM, respectively. The incident light $E_{in} = E_0 e^{j\omega t}$ has an amplitude and angular frequency of $E_0$ and $\omega$, respectively. Partial energy ($a(t)$) of the incident light enters the ring resonator with the mutual parameter $\mu$ and is stored inside the ring resonator with the behavior described by differential equation (1). This dynamic behavior is measured by the resonance wavelength of the Si-MRM $\omega_{res} = 2\pi Lmc/n_{eff}$ and the power decay rate constant $1/\tau = 1/\tau_e + 1/\tau_l$. Here, $m$, $c$ and $L$ are the resonance mode index, velocity of light in a vacuum, and ring circumference, respectively. $n_{eff}$, $\tau_e$ and $\tau_l$ are three voltage-dependent electrooptical parameters representing the effective refraction index of the ring waveguide, delay time constant due to ring resonance loss and coupling loss between the ring and bus waveguides, respectively. They have a nonlinear correlation with the voltage [36]. The mutual parameter can be related to $\tau_e$ by $\mu = \sqrt{2/\tau_e}$. For numerical solutions, the stored energy $a(t)$ in the $k_{th}$ time slot can be approximated by the following iteration equation [37]:

$$a_k(t) = \underbrace{\left[a_{k-1}(t_k) - Q_k(V_k)\exp(j\omega t_k)\right]\exp\left[\left(j\omega_{res,k}(V_k) - \frac{1}{\tau_k(V_k)}\right)(t-t_k)\right]}_{D_k(t,V_k)} + \underbrace{Q_k(V_k)\exp(j\omega t)}_{S_k(t,V_k)} \tag{3}$$

where $Q_k = -j\sqrt{2/\tau_e(V_k)}/\left[j(\omega - \omega_{res}(V_k)) + 1/\tau(V_k)\right]$. $V_k$ is the input voltage on Si-MRM in the $k_{th}$ time slot. $S(t,V_k)$ is the static solution at $t_{k-1} < t < t_k$ when the input voltage $V_k$ is in a steady state. $D(t,V_k)$ is the transient solution at $t_{k-1} < t < t_k$ that describes how the ring resonator responds to the variation in input voltage $V_k$. The sum of two solutions constitutes the overall dynamic electrical modulation behavior of the ring resonator. The final output optical power is subsequently calculated by equation (2), $O(t) = |E_{in} - j\mu a(t)|^2$.

Using equation (2-3), the modulation nonlinearity of Si-MRM is analyzed by observing the response of the Si-MRM to a multiple-level digital signal as shown in Fig.2. a. A digital signal

$O(t) = |E_{in} - jua(t)|^2$. The first one is the static nonlinear impairment resulting from the static solution $S(t,V_k)$. In the expansion equation of $S(t,V_k)$, the $S(t,V_k)$, $Q_k$ and the electrooptical parameters of Si-MRM have a nonlinear correlation with $Q_k$, the electrooptical parameters and $V_k$, respectively. The nonlinearity nesting causes the final complex nonlinear correlation of output optical signal with the input voltage signal (seen in the Fig.1b of Supplementary Note 1). The static modulation nonlinearity makes the steady-state signal-level interval unequal in $O(t)$, but it was originally identical in $V(t)$.

The second kind of nonlinear impairment is the overshoot phenomenon appearing on the rising and falling edges of $O(t)$. We define the overshoot phenomenon as transient nonlinear impairment since it originates from $D(t,V)$, the transient solution. The overshoot happens when $|D(t)|$ has a nonzero value and the value of $|D(t)|$ controls the amplitude of the overshoot. The transient nonlinear impairment degrades the level decision accuracy when the amplitude of the overshoot ($\Delta p$) is high enough to approach other levels. For example, the peak amplitude of level "-3" exceeds level "-5" in $O(t)$ due to the interference of overshoot. Hence, modulation formats with multiple levels, including the DMT modulation format, suffers from extra waveform distortion due to the transient nonlinear response besides the static nonlinear impairment. The transient nonlinearity causes the response of $a_k(t)$ has a time-relevant pattern dependency on the current signal level $V_k$ and the previous signal levels from the term $a_{k-1}(t)$. The larger wavelength detuning that the current and previous signal levels lead to, the more serious the overshoot becomes. The relation of wavelength detuning and overshoot has been validated in Supplementary Note 1. Fig.2 also illustrates this relation. The signal level with higher reverse amplitude will redshift the resonance wavelength of Si-MRM, leading to a larger wavelength detuning. These levels in $O(t)$ indeed have a bigger overshoot. For example, the overshoot of level "-5" in $O(t)$ is the biggest, inversely that of level "-1" is small. Besides, the overshoots in level "0" for different time slots have different amplitude though their signal levels are identical. The difference in overshoot attributes to the different signal levels at their last time slot, which verifies the relevance of the overshoot to the wavelength detuning of previous signal levels. Compared to level "0" whose last signal level is "-1" and "-3", the overshoots at the falling edge of level "-5" are highest because level "-5" has the largest wavelength detuning. More analysis of the time-relevant pattern dependency on signal levels for transient nonlinear impairments are seen in supplementary Note 1.

The different signal levels also change the optical bandwidth of Si-MRM. The optical bandwidth of Si-MRM can be approximated by the reciprocal of cavity photon lifetime ($1/\tau_p = \frac{\omega_{res}}{Q} = \frac{\omega_{res}}{\omega_{res}\tau/2} = \frac{2}{\tau}$). Since $\tau$ is sensitive to the signal level, the optical bandwidth varies according to the signal levels, causing different rising and falling time when the signal level shifts. The signal level with larger wavelength detuning has a larger optical bandwidth seen in the Supplementary Note 1.

Overall, when a high-speed electrical signal is modulated in the Si-MRM, the signal shift is expected to arouse the immediate response of the optical energy in ring resonator. Unfortunately, the variation is not abrupt due to the resonance nature of the resonator. A period of response time and oscillation accordingly appears before a new and steady resonance state is again reached, which causing the transient modulation nonlinearity. The response and the severity of oscillation depend on the synergetic wavelength detuning in current and previous electrical signal levels. The steady output power also has a static nonlinearity with the electrical due to the Lorentz-distribution filtering spectrum of Si-MRM.

*Demonstration of the 302Gbps optical interconnection using the Si-MRM and a Bi-GRU neural network*

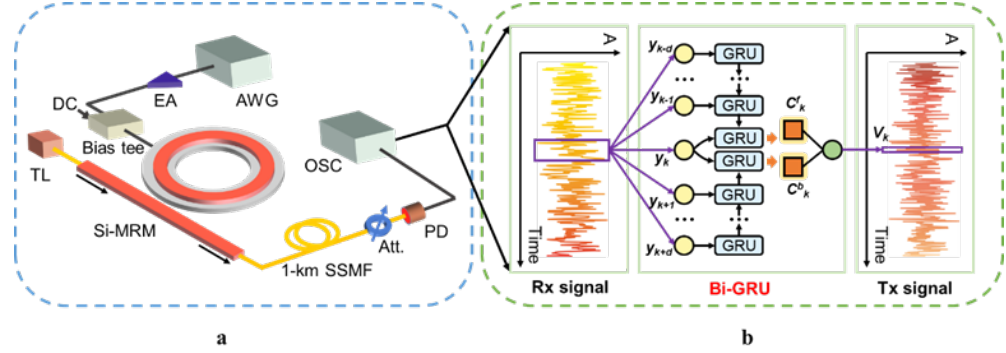

**Fig. 3 The proposed AI-accelerated Si-MRM-based optical transmission framework.** a. The proof-of-concept experimental system. (AWG: arbitrary waveform generator; EA: electrical amplifier; DC: direct current; TSL: tunable semiconductor laser; SSMF: standard single-mode fiber; Att.: attenuator; PD: photodetector; OSC: oscilloscope). b. The proposed Bi-GRU neural network used to reconstruct the transmitted signal (Tx signal) from the received signal (Rx signal). (A: Amplitude).

Based on our new insights into the nonlinear modulation characteristics of Si-MRM, we achieve an ultra-high-speed Si-MRM-based optical interconnection with AI acceleration. The experimental setup is shown in Fig. 3a with the details in Methods. To maximize the SE and approach the Shannon limit, the BPL-DMT modulation is applied in this experiment to allocate bits in frequency subcarriers adapting the uneven channel. This modulation format has been demonstrated more spectral-efficient than single carrier modulation, such as pulse amplitude modulation for the uneven channel [38]. The digital signal processing (DSP) flow chart of the BPL-DMT signal generation and recovery in this experiment is seen in Supplementary Note 2 and Methods. A modified Bi-GRU neural network is used to reconstruct the transmitted signal V(t) from the received signal y(t) suffering from linear and nonlinear impairments before the normal BPL-DMT demodulating process. The Bi-GRU neural network consists of three layers: an input layer, a hidden layer and an output layer. The hidden layer is composed of two GRU cells with different time-flow directions. Instead of using the identical received signal samples for two directional GRU cells in traditional Bi-GRU [39], the data entering the forward GRU is the received signal from $(k-d)_{th}$ to $k_{th}$ time slots, and the data entering the backward GRU is the received signal from $k_{th}$ to $(k+d)_{th}$ time slots. The overall time window is $(2d+1)$. Their final memory state vectors $C_k^f$ and $C_k^b$ from the forward and backward GRUs respectively, cascade together into one vector connecting to the output layer with only one node.

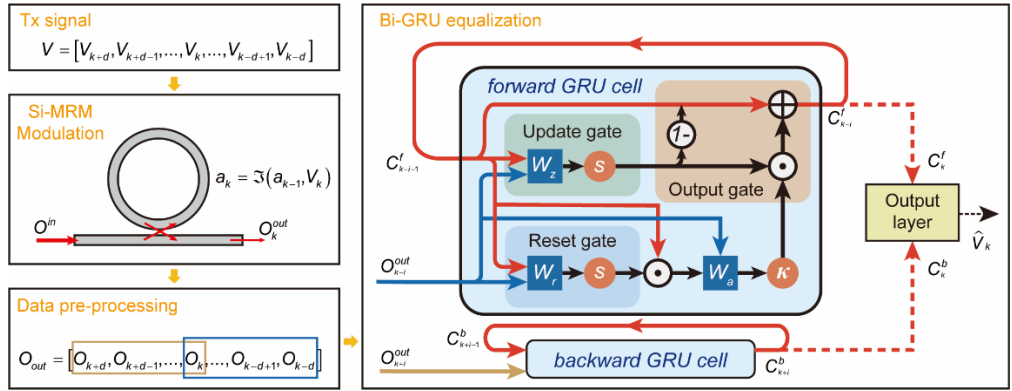

**Fig. 4 The principle of Bi-GRU serving as a nonlinear equalizer and the inner configuration of a GRU cell.**

Rather than a "black box" application of AI merely driven by hundreds of samples, the inner configuration of the Bi-GRU neural network matches the dynamic characteristics of Si-MRM. The feasibility of Bi-GRU serving as an effective nonlinear equalizer is analyzed by comparing its inner configuration with dynamic characteristics of Si-MRM illustrated in Fig.4. According to our abovementioned theoretical analysis, the transient modulation nonlinearity makes the stored energy in the ring resonance cavity ($a_k$) has a time-relevant pattern dependency on current signal level ($V_k$) and previous response ($a_{k-1}$), where k represents one certain time slot. The time-relevant pattern dependency can be simplified to $a_k = \Im(a_{k-1}, V_k)$, where $\Im(\cdot)$ stands for a complex nonlinear time-relevant correlation presented in equation. (3). Since the output optical signal $O_k^{out}$ satisfies a simple square relation with the stored energy ($O_k^{out} = |E_{in} - jua_k|^2$), the output optical signal also has a time-relevant pattern dependency written as $O_k^{out} = \Re(O_{k-1}^{out}, V_k)$. Inspired by the equalization mechanism of the traditional Volterra filter [40], an equalizer having a similar model with the physic system can approximate the inversion of the system by the fixed point approach. Fortunately, the forward GRU cell happens to provide the time-relevant model that the memory state from last time slot and input sample (i.e. the received signal from the oscilloscope) contribute to current memory state, expressed as $C_{k-i}^f = \aleph(C_{k-i-1}^f, y_{k-i}^{out})(0 \leq i \leq d)$, where the inner mathematic model $\aleph(\cdot)$ is composed of three gates, update gate, reset gate and output gate. The Bi-GRU's received signal y(t) is directly simplified to the output of Si-MRM O(t) here, neglecting the noise, and thus $C_{k-i}^f = \aleph(C_{k-i-1}^f, O_{k-i}^{out})(0 \leq i \leq d)$. The three gates have nonlinear activation functions ($s$: sigmoid function, $\kappa$: tanh function) and trainable parameters ($W_z$, $W_r$ and $W_a$) to find the optimal numerical solution of the Si-MRM equalization problem. Meanwhile, the inter-symbol interference (ISI) effect also makes the current symbol contain symbols in future time slots. The ISI effect comes from the bandwidth limitation of Si-MRM and other E-O devices. Hence, a backward GRU cell fed into the subsequent symbols relative to the $k_{th}$ time slot is also necessary to help reconstruct the transmitted symbol in the $k_{th}$ time slot. After the chain propagation of the memory state in Bi-GRU, i.e. the value of i in $C_{k-i}^f$ and $C_{k+i}^f$ decrease from d to 0, the final outputs of two GRU cells ($C_k^f$ and $C_k^b$) calculates the equalized symbol $\hat{V}_k$ through the output layer. The specific calculation process in the GRU cell is given in Ref. [41] and Supplementary Note 3. The Bi-GRU has better equalization performance than data-driven fully-connected neural network equalizers and is robust for different signal patterns thanks to its superior structure, which is further analyzed in Supplementary Note 3.

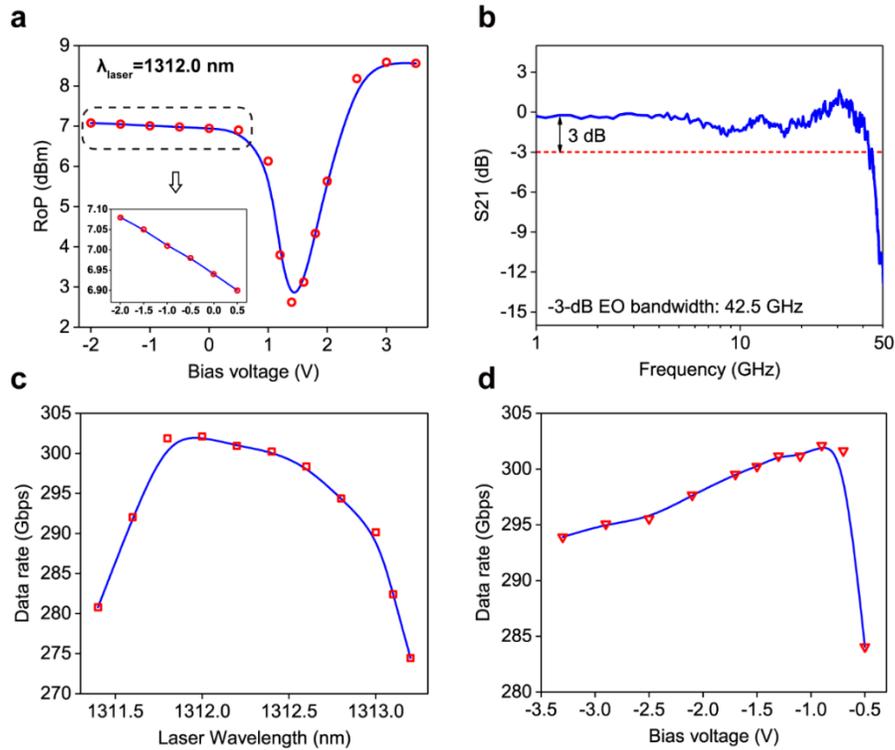

**Fig. 5 Modulation characteristics of the Si-MRM.** a. E-O transfer curve of Si-MRM when the laser wavelength is 1312.0 nm. Inset: the amplified picture of the E-O transfer curve at the range of -2 V to -0.5 V. b. System frequency response of our experimental optical link based on the Si-MRM. c. Data rate versus different laser wavelengths. d. Data rate versus different bias voltages. The Vpp at c and d are both fixed at 600 mV.

The modulation characteristics of our proposed Si-MRM are shown in Fig. 5. The E-O transfer curve of our Si-MRM at 1312 nm in Fig.5a is obtained by measuring the Rops at different bias voltages. The Rop seriously decreases when the bias voltage approaches the voltage (~1.5 V) where the laser wavelength equals the resonance wavelength. The two sides of the E-O transfer curve have different RoPs because Si-MRM at negative bias voltage is on the depletion mode but the carrier injection mode at positive bias voltages. The attenuation of Si-MRM at the two modes is different. To get a wider E-O bandwidth, the bias voltage needs to be a smaller and negative value with a larger wavelength detuning according to our above analysis of the dynamic characteristics of Si-MRM. The channel response at -0.9 V is presented in Fig. 5b, in which the -3-dB E-O bandwidth is approximately 42.5 GHz and the overall frequency response is uneven and suffers from serious fading issues. The peaking appears at 30 GHz.

Though the E-O bandwidth can be extended by decreasing the bias voltage or the laser wavelength, the extinction ratio, regarded as the slope of the E-O transfer curve, accordingly decreases indicated in Fig.5a. In addition, the induced increase of wavelength detuning intensifies the transient nonlinear impairment. The bias voltage of -0.9 V and laser wavelength of 1312 nm could achieve the highest data rate by iteration optimization. As shown in Fig.5c, when the laser wavelength increases, the wavelength detuning decreases with a larger extinction ratio, and thus the achievable data rate rises due to improved signal-to-noise ratio (SNR), although the bandwidth gradually decreases. When the laser wavelength is redshifted over 1312 nm, the signal swing reaches the steep slope of the E-O transfer curve. The achievable data rate thus inversely decreases due to decreased Rop and more serious static nonlinear impairment which all decrease the SNR. In the same vein, as the bias voltage moves from -3.2 V to -0.9 V, the increase of extinction ratio improves SNR and data rate. The data rate declines when the bias voltage approaches -0.5 V since the decline of Rop and serious static nonlinear impairment. The static nonlinear impairment causes asymmetric amplitude distribution in the received optical signal. From the measured amplitude ratio (AM/AM) of the transmitted electrical signal and the received electrical signal in the back-to-back case shown in Fig.6a, if the transmitted signal amplitude is normalized to [-1, 1], the response of [0,1] is much larger than that of [-1,0] where the amplitude of the optical signal is limited to [-0.5, 1]. The distribution of the received electrical signal is Lorenz distribution with a similar shape to the E-O transfer curve after cubic fitting, validating the existence of the static nonlinear impairment with 3-orders.

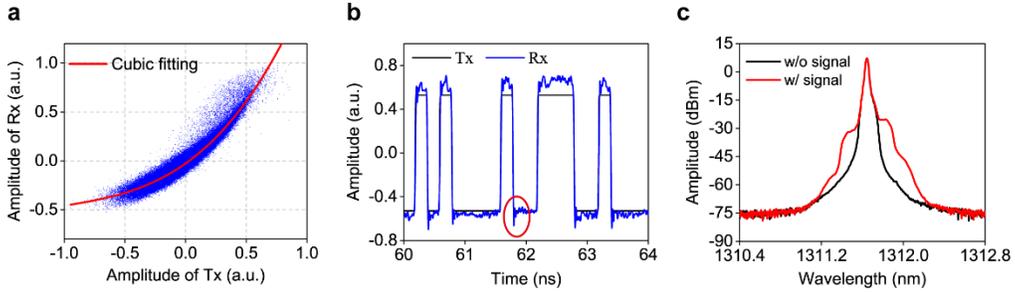

**Fig. 6 Results of the nonlinear impairment analysis.** a. The measured AM/AM curve of the time-domain transmitted (Tx) and received (Rx) signals for the MRM at the laser wavelength of 1312 nm, bias voltage of -0.9 V and Vpp of 0.6 V. b. The waveform of the Tx and Rx signal after 5 GBaud NRZ modulation. The red circle marks the overshoot phenomenon, i.e. the transient nonlinear impairment. c. Optical spectrum of the light output from the Si-MRM with/without the modulation of a 60 GBaud (302 Gbit/s) DMT signal.

Fig. 6b shows the waveform of the received electrical NRZ signal and the transmitted NRZ signal. The baud rate and Vpp of the signal are 5 GBaud and 600 mV, respectively, and modulated at the optimal working point. The wavelength detuning of the negative signal level is much larger than that of the positive signal, causing the overshoot in the negative level marked by a red circle. On the contrary, the resonance wavelength of the positive signal level is closed to the laser wavelength without showing evident overshoot. The experimental results are consistent with the simulation results in Fig.2 that overshoot amplitude is related to the wavelength detuning. Analogously, the transient nonlinear impairment will impose transient nonlinear distortion on the DMT signal that has multiple levels. But, it is uneasy to directly observe the distortion from waveform due to the irregular signal level distribution as a kind of analog signal. The optical spectrum of the DMT signal could indirectly validate the distortion of transient nonlinear impairment on the waveform. The signal suffering from the transient nonlinear impairment will exhibit an asymmetric spectral shape due to the chirp effect [42]. Fig.6c shows the optical spectrum of the 302 Gbps BPL-DMT signal and that of the optical carrier. It presents distinct asymmetry in the left and right sidebands. In addition, the peak wavelength blueshifts to 1311.6 nm due to the wavelength conversion effect in Si-MRM [43].

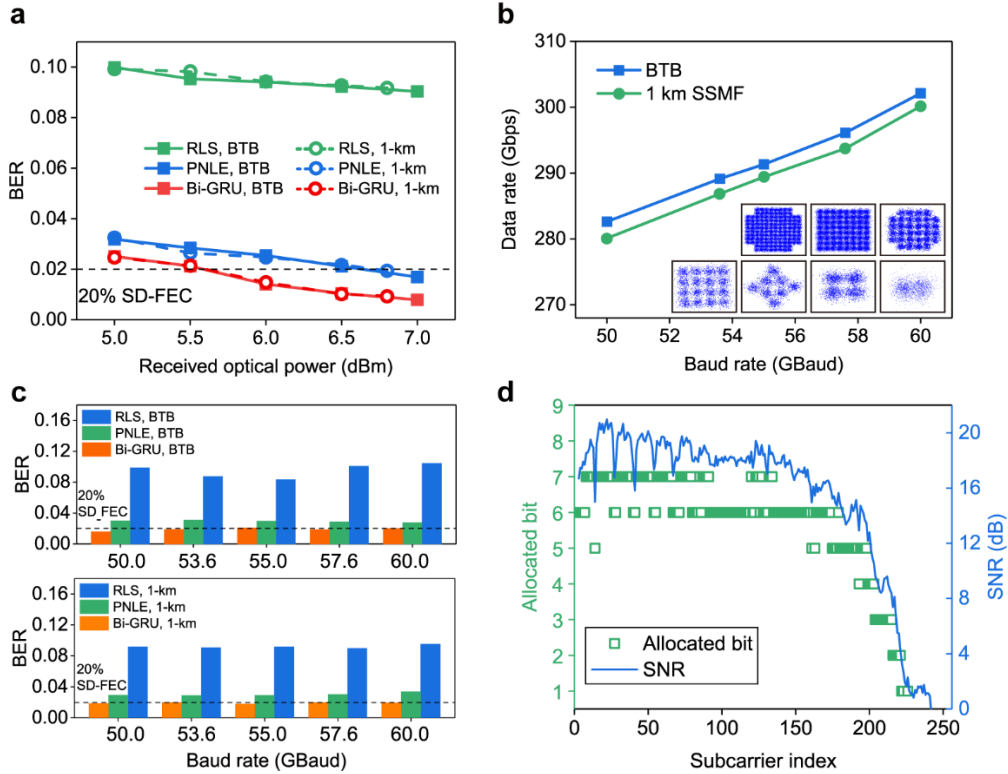

**Fig. 7 Experimental results of the proof-of-concept Si-MRM-based optical interconnection.** (a) BER performance of three equalization schemes versus different received optical powers at a data rate of 280 Gbps. (RLS: FFE based on recursive least squares algorithm. PNLE: 3rd-order polynomial nonlinear equalizer). (b) Data rate versus baud rate in the cases of back-to-back (BTB) and 1 km SSMF. Insets: constellation diagrams of the received QAM signal at a data rate of 302 Gbps. (c) BER performance of three equalization schemes versus different baud rates for the BTB and 1km fiber transmission cases. (d) The SNR distribution and bit allocation at different subcarriers when the data rate reaches 302 Gbps.

Next, a high-speed transmission experiment based on the Si-MRM and the Bi-GRU neural network is carried out. We first evaluate the BER performance of the Bi-GRU neural network compared to that of other traditional linear and nonlinear equalizers at different RoPs. The preset data rate of the transmitted BPL-DMT-modulated signal is 280 Gbps with a baud rate of 60 GBaud. The linear equalizer utilized for comparison is a feedback forward equalizer (FFE) based on the recursive least squares (RLS) algorithm. The nonlinear equalizer is the third-order polynomial nonlinear equalizer (PNLE) based on the RLS algorithm, which has sufficient equalization ability for third-order static nonlinear impairment. The results in Fig. 7a show that both the PNLE and Bi-GRU greatly outperform the linear RLS equalizer, which means that nonlinear distortion occupies a significant part of the whole signal distortion. Compared to the PNLE, the Bi-GRU provides an ~1.1-dB gain of the RoP under the 20% SD-FEC threshold because the Bi-GRU considers the extra transient nonlinear impairments and has better optimization ability. The BER performance for transmitting over 1 km shows a similar level to that of the BTB case, which verifies the ability of the Si-MRM for high-speed short-reach optical interconnection. To obtain the maximum data rate achievable by the Si-MRM, we test the data rate under different baud rates when the RoP is set to 7 dBm (Fig. 7b). The data rate at the BTB could reach 302 Gbps at a baud rate of 60 GBaud after removing the redundancy of 8 zero-padded subcarriers with an SE of ~5.20 bit/s/Hz ( $data\ rate = \frac{SE \times B \times (Fs - 2z)}{Fs}$ ). The $B$, $Fs$ and z present baud rate, FFT size and the number of zero-padded subcarriers, respectively. The maximum data rate over the 1-km SSMF is 300 Gbps with an SE of ~5.16 bit/s/Hz. The BER performance of the three equalization schemes shown in Fig. 7c further verifies the powerful resistance of the Bi-GRU to the linear and nonlinear impairments existing in the Si-MRM. Only the Bi-GRU could decrease the BER to below 2e-2 to help achieve 302 Gbps. Fig. 7d shows the SNR and bit distribution of the DMT signal at a data rate of 302 Gbps. There are 256 subcarriers distributed at the 60 GHz baseband, in which the first two subcarriers and last six subcarriers are abandoned. With the aid of the Bi-GRU and the bit and power loading technology, bits were adaptively allocated to every subcarrier. The low-frequency subcarriers could support 32-QAM to 128-QAM, but only BPSK could transmit at the subcarriers around 50 GHz, i.e. around the 220$^{th}$ subcarriers. The constellation diagrams from BPSK to 128-QAM for the 302 Gbps data rate are presented as insets in Fig. 7b.

**Discussion**

To improve the Si-MRM performances, we comprehensively optimize the Si-MRM from the device, to modulation as well as the signal processing. We fabricate a racetrack Si-MRM with a -3-dB optical bandwidth over 67 GHz enabled by doping concentration optimization and OPE technology. The modulation characteristics of the Si-MRM is experimentally studied. Consequently, the laser wavelength and bias voltage of Si-MRM is optimized to improve the modulation performance. Although the bandwidth is greatly extended, nonlinear impairments induced by the Si-MRM dramatically restrict the overall modulation capacity. To fully compensate for this, we theoretically analyze the mechanism of two kinds of nonlinear impairments and their dependency on the signal level. This analysis helps to fully understand the modulation nonlinearity in Si-MRM, and establishes the connection between the physical model and advanced signal processing technologies such as AI. Consequently, a Bi-GRU neural network with minor modification based on the obtained physical model of Si-MRM's modulation nonlinearity is applied to help accelerate the Si-MRM modulation speed. We experimentally demonstrate that by selecting a suitable configuration of neural networks according to the dynamic nonlinear response of the Si-MRM, the AI technology can accelerate the modulation of the Si-MRM to a record-breaking data rate of 302 Gbps. The Bi-GRU neural network presents a better equalization performance than traditional nonlinear equalizer. The results in this work offer deeper insight into DSP for Si-MRM-based high-speed optical interconnections, validating the concept that the AI technologies based on the deep understanding of physics system could further improve the modulation capacity of Si-MRMs. The concept could also be promoted to other disciplines and facilitate their development.

**Methods**

*Methods for parameter extraction of the Si-MRM*

The key step in simulating the Si-MRM is to obtain the correlation of $n_{eff}$, $\tau_e$ and $\tau_l$ with the input voltage. They can be extracted by using the high-order polynomial expansion to fit the transmission spectra of the Si-MRM at different input voltages. The transmission spectrum

$$T(V) = \left| \frac{j(\omega - \omega_{res}(V)) + \frac{1}{\tau_l(V)} - \frac{1}{\tau_e(V)}}{j(\omega - \omega_{res}(V)) + \frac{1}{\tau_l(V)} + \frac{1}{\tau_e(V)}} \right|^2$$ [44]. The transmission spectrum is determined by the structure and

doping concentration of the P-N junction in the Si-MRM [36]. In our simulation, the ring radius, doping concentration of N, doping concentration of P and junction feature length are 10 μm, 8.2e24 (1/cm$^3$), 3.5e24 (1/cm$^3$) and 200 nm, respectively. By fitting the transmission spectrum, the correlations of $n_{eff}$, $\tau_e$ and $\tau_l$ with the input voltages are:

$$n_{eff} = -1.3686e^{-10}V^4 - 3.4678e^{-9}V^3 - 3.6345e^{-8}V^2 - 3.1326e^{-7}V + 0.0208 \quad (4)$$

$$\tau_e = -2.661e^{-15}V^3 - 5.7557e^{-14}V^2 - 1.3669e^{-12}V + 6.2055e^{-11} \tag{5}$$

$$\tau_l = 4.1324e^{-15}V^3 + 9.1925e^{-14}V^2 + 9.2863e^{-12}V + 3.4369e^{-11} \tag{6}$$

*Experimental setup of the optical interconnection*

The arbitrary waveform generator (AWG, Keysight M8194A) first generates the signal stream to be transmitted, which is then amplified by an electrical amplifier (SHF S807C, 55 GHz). Next, a wide-bandwidth bias tee (SHF BT65B) imposes the amplified signal on a bias voltage to drive the P-N junction of the Si-MRM by a high-frequency ground-signal radio-frequency probe. The bias voltage is supplied by a direct current (DC) power supply (KEITHLEY 2230G-30-1). The optical carrier is provided by a tunable semiconductor laser (TSL, Santec TSL-550) with a laser wavelength of 1312.0 nm and coupled into the bus waveguide through a fiber patch cord and a grating coupler. The coupling efficiency for the transverse electric (TE) mode increases to 40% with the grating coupler, which is also used in the distal end of the bus waveguide. The output optical signal is converted to an electrical signal by a photodetector (PD, FINISAR XPDV2320R) after transmitting over a 1-km SSMF and an optical attenuator (Att.). The electrical signal is captured by an oscilloscope (KEYSIGHT UXR0704A), which has a maximum sample rate of 256 GS/s.

*DSP of BPL-DMT signal generation and recovery*

Supplementary Fig. 2 presents the algorithmic procedure of the BPL-DMT, which consists of two steps. The first step is to estimate the SNR at every subcarrier of the DMT signal. Then, the estimated SNR guides the second step, i.e., bit and power loading, to adaptively allocate the optimal bit and power on every subcarrier. In the first step, we generate a random pseudorandom bit sequence (PRBS) based on the Mersenne-twister algorithm [45] by using MATLAB, which is then mapped to a QAM-4 signal serial sequence. After being converted to a parallel sequence, the QAM-4 signal is modulated to a DMT signal with a fast Fourier transform (FFT) size of 512 according to the standard DMT modulation process [46]. The cyclic prefix (CP), with a length of 16, is added to the head of every DMT symbol to eliminate the ISI. The parallel DMT signal is then transferred to a serial DMT signal, which is the ultimate training signal fed into the AWG. The maximum amplitude of the DMT signal is normalized to a unit value. Considering the poor response at low and high frequencies of the channel, the first two subcarriers and last six subcarriers are unused here. At the receiver, the captured signal is first synchronized and resampled to the same baud rate as that of the transmitted signal. Then, the Bi-GRU is used to equalize the linear and nonlinear impairments of the signal. After that, the equalized signal is transferred to the frequency domain by the inverse FFT operation. The residual linear distortion of each subcarrier on the signal is equalized by the zero-forcing equalizer (ZFE) and intrasymbol frequency-domain averaging (ISFA) technologies. The sampling frequency offset (SFO) is mitigated using two DMT symbol pilots that lie on the head and tail of the DMT signal sequence introduced in Ref. [47]. Subsequently, the error-vector magnitude of the signal is used to estimate the SNR of every subcarrier under the preset 20% SD-FEC BER threshold of 2e-2 [48, 49]. It produces an optimal QAM order and power ratio allocation strategy for each subcarrier based on the Levin-Campello algorithm [50, 51]. In the second step, another group of PRBSs is modulated to a new DMT signal sequence based on the allocation strategy obtained in the first step. The modulation and demodulation process and DSP process are the same as those in the first step. The BER is calculated by using the recovered PRBS and the transmitted PRBS.

*Parameters of the Bi-GRU neural network*

The ratios of the training set, validation set and test set are 30%, 20% and 50%, respectively. To achieve the best equalization performance and decrease the parameter complexity as much as possible, the optimal length of time windows and optimal dimensions of the memory state are 11 and 80, respectively, obtained by parameter iteration optimization. The batch size and training epoch are set to 64 and 60, respectively, to reduce the amount of time spent on back-propagation.

## Data availability

The data files are available from the corresponding author upon reasonable request.

## Code availability

Example codes for the training and testing of the deep neural networks used in the paper are available from the corresponding author upon reasonable request.

## Supplementary Information

Supplementary information is provided in https://drive.google.com/file/d/1icsSdtojD-BjdAQBmoHG8LiDKgT2kNHR/view?usp=sharing

## Acknowledgments

This work was partly support by the NSFC project (No.61925104, No.62031011, No. 62171137) Shanghai NSF project (No. 21ZR1408700) and The Major Key Project of PCL (PCL2021A14).

## Author contributions

FCH, YGZ and HGZ contributed equally to this work. FCH, HGZ, YGZ, ZYL, SZX and JWZ conducted the experiments. FCH, ZYL, JYS, NC and JWZ contributed to the theoretical analysis and simulation. FCH analyzed the data and wrote the paper. YGZ, HGZ and XX fabricate the Si-MRM and establish the experimental setup. ZXH, NC and SHY supervised the research project. All authors commented on the manuscript.

## Competing interests

The authors declare no competing interests.